\documentclass[reprint,prb]{revtex4-1}
\usepackage{bm,epsfig,graphicx,color,amssymb,amsmath}

\begin{document}

\title{Long-range plasmon-assisted energy transfer over doped graphene}

\author{Kirill A. Velizhanin}

\email{kirill@lanl.gov}

\affiliation{Theoretical Division, Los Alamos National Laboratory, Los Alamos,
New Mexico 87545, USA}

\author{Tigran V. Shahbazyan}

\email{shahbazyan@jsums.edu}

\affiliation{Department of Physics, Jackson State University, Jackson, Mississippi
39217, USA}

\keywords{resonance energy transfer, graphene, plasmons, semiconductor quantum dots}

\begin{abstract}
We demonstrate that longitudinal plasmons in doped monolayer graphene can mediate highly efficient long-range energy transfer between nearby fluorophores, e.g., semiconductor quantum dots. We derive a simple analytical expression for the energy transfer efficiency that incorporates all the essential processes involved. We perform numerical calculations of the transfer efficiency for a pair of PbSe quantum dots near graphene for inter-fluorophore distances of up to 1 $\mu$m and find that the plasmon-assisted long-range energy transfer can be enhanced by up to a factor of $\sim$10$^4$ relative to the F\"{o}rster's transfer in vacuum.
\end{abstract}

\maketitle

\section{Introduction}

F\"orster resonance energy transfer (FRET) \cite{Forster1948-55} between spatially separated donor and acceptor fluorophores, such as dye molecules or semiconductors quantum dots (QD), underpins diverse phenomena in physics, chemistry and biology. Examples include photosynthesis, exciton transfer in molecular aggregates, interactions between proteins \cite{Lakowicz2006,Andrews1999} and, more recently, energy transfer between QDs and QD-protein assemblies.\cite{Willard2001-469,Crooker2002-18,Clark2007-7302}
During the past decade, remarkable progress has been made in applications of FRET spectroscopy, e.g.,  in protein folding,\cite{Deniz2000-5179,Lipman2003-1233} live cell protein localization,\cite{Selvin2000-730,Sekar2003-629} biosensing \cite{Gonzales1995-1272,Medintz2003-630} and light harvesting.\cite{Andrews2011-114} The range of  present and potential applications of FRET is, however, limited by its intrinsically short-range nature. Indeed, the underlying FRET mechanism -- the direct Coulomb interaction between fluorophores -- supports efficient transfer only at donor-acceptor distances ($r_{ad}$) below the typical F\"orster radius of $r_{F}\sim10$ nm.\cite{Lakowicz2006} At larger distances, the Coulomb potential  between electrically neutral donor and acceptor decreases rapidly, and the FRET efficiency falls off as  $\sim$$r^6_F/r^6_{ad}$. Substantial efforts have been undertaken to improve the {\em efficiency} and increase the {\em range} of energy transfer (ET)  at the nanoscale by utilizing surface plasmons (SP) and surface plasmon-polaritons (SPP) as intermediaries.\cite{Andrew2004-1002,Lakowicz2003-69,Lakowicz2007-50,Reil2008-4128,Komarala2008-123102,Yang2009-11495,An2010-4041}  Placing molecules or QDs near a metal film or a nanoparticle can lead to a significant improvement of ET efficiency (ETE) -- the fraction of donor's energy 
transferred to the acceptor.\cite{Gersten1984-31,Hua1985-3650,Druger1987-2649,Dung2002-043813,Durach2008-105011,Pustovit2011-085427} In metals, however, the efficiency of plasmon-mediated ET channels is  limited by significant Ohmic losses and  plasmon-enhanced radiative losses \cite{Pustovit2011-085427} resulting in a relatively modest ($\sim$10) overall ETE increase \cite{Andrew2004-1002,Lakowicz2003-69,Lakowicz2007-50,Komarala2008-123102,Yang2009-11495,An2010-4041}
or even its reduction \cite{Leitner1988-320,Reil2008-4128} near metal structures.
\begin{figure}
\includegraphics[width=1.0\columnwidth]{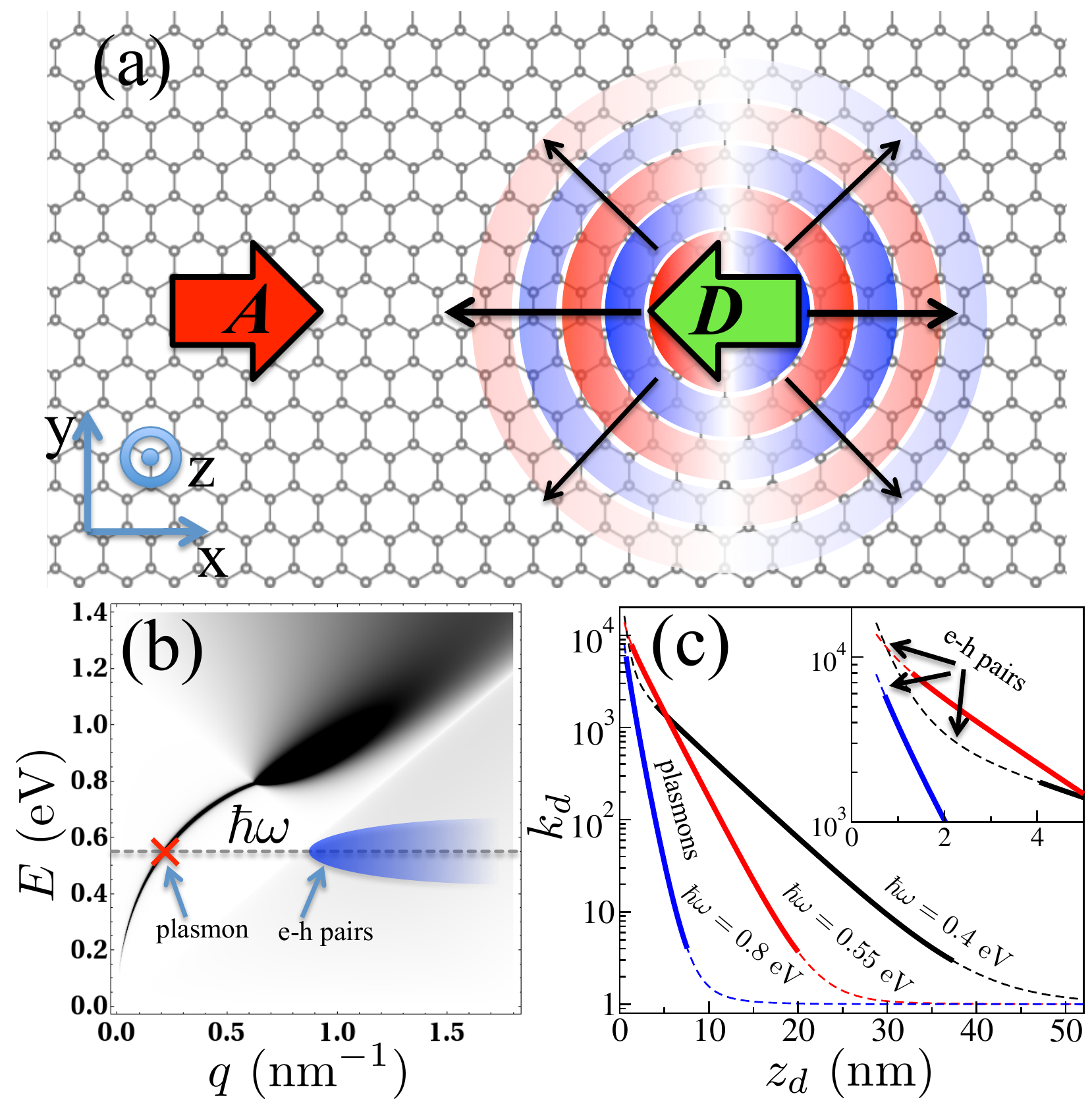} \caption{\label{fig:schematic} (a) Schematic of ET between donor (D) and acceptor (A) through the GP excitation. (b) Density plot of the imaginary part of graphene density correlation function (darker shades of gray correspond to higher magnitudes). The Fermi level is set to $\epsilon_f=0.6$~eV. (c) Normalized donor energy loss rate versus donor-graphene distance for suspended graphene in vacuum. The energy loss rate is plotted for three different excitation energies. Thick solid portions of lines mark the intervals of the exponential decay of $k_d$ -- intervals where GP is most efficiently excited (tagged ``plasmon" in the graph). The inset shows the zoomed-in left-top portion of the main graph in the panel (c). Intervals of efficient electron-hole pair excitation at low $z_d$ are tagged ``e-h pairs" in the inset.}
\end{figure}

In this article, we propose to exploit collective excitations in graphene as efficient ET intermediaries.
Graphene has recently emerged as a novel intrinsically two-dimensional material \cite{Novoselov2004-666,Geim2007-183} with unique electronic and optical properties.\cite{CastroNeto2009-109} Clean graphene samples are characterized by long electron scattering times and much lower, compared to metals, Ohmic losses due to relatively weak electron-phonon interaction.\cite{Xiao2007-206} Furthermore, doped graphene possesses a stable in-plane longitudinal plasmon in the infrared frequency range with gate-tunable wavelength, $\lambda_{p}$, well below radiation (or SPP)  wavelength $\lambda_{0}$ at the same frequency.\cite{Jablan2009-245435} Recent optical imaging of graphene plasmons (GP) propagating in a graphene ribbon on SiC substrate indeed demonstrated the high degree of GP localization characterized by light-to-GP wavelength ratio of $\lambda_{0}/\lambda_{p}\approx40$.\cite{Fei2012-82,Chen2012-77} The large GP local density of states, as compared to that of SPP, permits very efficient GP excitation by a local 
probe such as a scanning tunneling microscope (or atomic force microscope) tip, an excited molecule or QD placed at a close distance ($\lesssim$$\lambda_{p}$) from the graphene sheet.\cite{Velizhanin2011-085401,Koppens2011-3370,Nikitin2011-195446} Recently, superradiance from two emitters placed near graphene was studied,\cite{Huidobro2012-155438} and it was demonstrated that the interaction of fluorophores with plasmons in graphene can be strong enough to significantly enhance the superradiant coupling between these fluorophores.

\begin{figure}
\includegraphics[width=1.0\columnwidth]{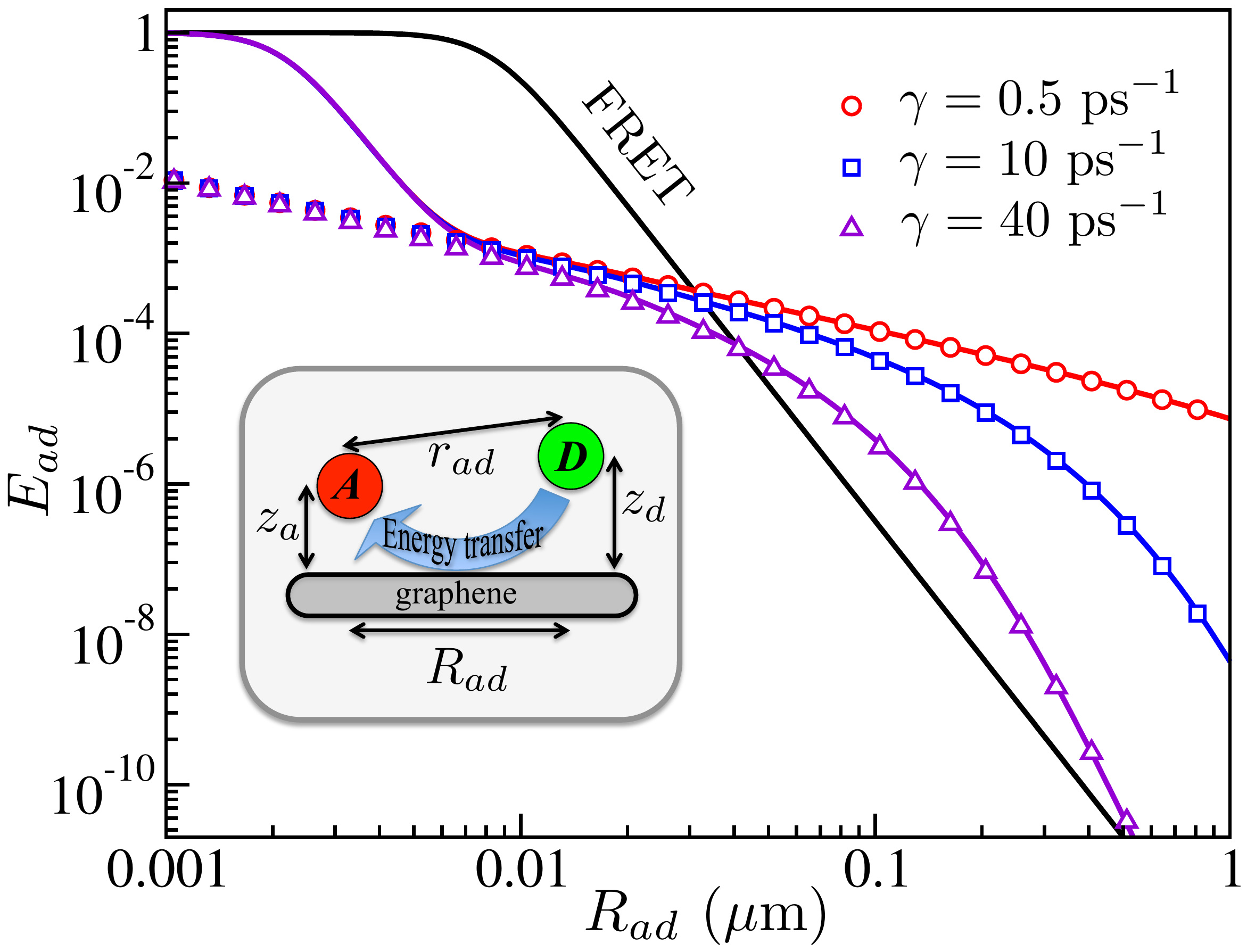} \caption{\label{fig:Ead_num_anl} ETE versus in-plane distance between donor and acceptor near suspended graphene in vacuum ($\tilde{\kappa}=1$). The Fermi level is set to $\epsilon_f=0.6$~eV. Analytical and numerical results are shown by symbols and solid lines, respectively. ETE with and without graphene is shown in color and black, respectively. Further details are provided in the text. 
}
\end{figure}

In this work, we demonstrate that plasmons in doped graphene can mediate a highly efficient \textit{long-range} ET between
fluorophores, e.g., QDs. A photoexcited donor with energy $\hbar\omega$ situated at a distance $z_{d}$ from the graphene
sheet excites a GP which propagates a distance $R_{ad}\gg\lambda_{p}$ in the plane before exciting a remote acceptor at
a distance $z_{a}$ from graphene (see schematics in Figure~\ref{fig:schematic} and Figure~\ref{fig:Ead_num_anl}). Importantly, when the GP wave reaches the acceptor, its intensity is reduced only by factor of $\propto$$\lambda_{p}/R_{ad}$ due to the strictly in-plane GP propagation.  This, along with the efficient fluorophore-GP coupling at $z_{a,d}\lesssim \lambda_{p}$, leads to a very strong ET enhancement  (up to $\sim$$10^{4}$) as compared to the FRET channel, at distances far exceeding the F\"orster radius. 

We show that at large transfer distances, $R_{ad}\gg\lambda_p$, ETE between donor and acceptor is given by (see Section~\ref{sec:Theory} for the detailed derivation)
\begin{equation}
E_{ad}^{gp}=D_{p}/R_{ad},\label{eq:ete_pl}
\end{equation}
where 
\begin{align}
D_{p}(R_{ad})&=\frac{4}{3\tilde{\kappa}}\int d\omega\, q_{p}^{2}(\omega)f_{d}(\omega)\alpha''_{a}(\omega)
\nonumber \\
&\times e^{-R_{ad}/R_{p}(\omega)-2q_{p}(\omega)|z_a|}\label{eq:distance_pl}
\end{align}
is the characteristic ET length which, in high mobility graphene, only weakly depends on $R_{ad}$. Here, $\alpha_{a}(\omega)=\alpha'_{a}(\omega)+i\alpha''_{a}(\omega)$ is acceptor's complex dipole polarizability, $q_{p}(\omega)$ and $R_{p}(\omega)$ are GP wavenumber and characteristic travel length, respectively; $f_{d}(\omega)$ is donor's normalized emission spectral function, and  $\tilde{\kappa}$ is the effective dielectric constant of the environment ($\tilde{\kappa}=1$ for vacuum and $\tilde{\kappa}=2.5$ for ${\rm SiO_2}$ substrate). 

Figure~\ref{fig:Ead_num_anl} shows our numerical and analytical results for  ETE between PbSe QDs
a near graphene sheet doped to the Fermi level of $\epsilon_{f}=0.6$~eV
for several values of electron scattering rate $\gamma$. Numerical results are obtained using the full graphene density
correlation function, while analytical results, given by Eqs.~(\ref{eq:ete_pl}) and (\ref{eq:distance_pl}), are obtained within the plasmon pole approximation (see Section~\ref{sec:Theory}); they are in excellent agreement for distances exceeding GP wavelength $\lambda_{p}\approx 30$ nm. The large-distance behavior of ETE depends strongly on the sample quality characterized by $\gamma$, which, in turn, determines GP travel length $R_p$. As is seen, for low-$\gamma$ samples, GP-assisted ETE exceeds F\"{o}rster's ETE in vacuum with $r_{F}=8 $ nm (shown by the solid black line) by a factor of $\sim$10$^2$ for $R_{ad}=100$ nm and by factor of $\sim$$10^{4}$ -- $10^{6}$ for $R_{ad}=500$ nm. Such extreme enhancements are due to the slow decay of two-dimensional GP wave amplitude. For high-$\gamma$ samples, ET is limited due to reduced GP travel length $R_p$.

In the rest of the paper, we derive Eqs.~(\ref{eq:ete_pl}) and (\ref{eq:distance_pl}), and study numerically and analytically ET over graphene for various system parameters.

\section{Theory}\label{sec:Theory}

We consider donor and acceptor fluorophores (dye molecules or QDs) as point dipoles situated at ${\bf r}_{i}=({\bf
R}_{i},z_{i})$, ($i=a,d$) with transition dipole moments $\bm{\mu}_{i}=\mu_{i}{\bf n}_{i}$, where ${\bf n}_{i}$ is the
dipole orientation, separated from each other by $r_{ad}=\left [R_{ad}^{2}+(z_a-z_d)^{2}\right ]^{1/2}$ (see inset in Figure~\ref{fig:Ead_num_anl}). If $z_{i}$ is not too small so that fluorophores' internal transitions are not significantly affected by graphene, ETE can be found within the semiclassical approach.\cite{Novotny2006,Pustovit2011-085427} The power transferred from the donor, initially excited at frequency $\omega$, to the acceptor is given by 
\begin{equation}
P_{ad}(\omega)=\frac{\omega}{2}\alpha''_{a}(\omega)\left|{\bf n}_{a}\cdot{\bf E}(\omega;{\bf r}_{a})\right|^{2},\label{power}
\end{equation}
where ${\bf E}(\omega;{\bf r}_{a})$ is the electric field at the acceptor's position. This field is related to the donor's dipole moment via 
\begin{equation}
{\bf E}(\omega;{\bf r})=\frac{4\pi\omega^{2}}{c^{2}}{\bf G}(\omega;{\bf r},{\bf r}_{d})\cdot\bm{\mu}_{d},\label{electric}
\end{equation}
where ${\bf G}(\omega;{\bf r},{\bf r}')={\bf G}_{0}(\omega;{\bf r},{\bf r}')+{\bf G}_{g}(\omega;{\bf r},{\bf r}')$
is the electric field Green dyadic comprised of direct and graphene-assisted contributions, respectively. For brevity,
we introduce a matrix $S_{ij}(\omega)=\left(4\pi\omega^{2}/c^{2}\right)\,{\bf n}_{i}\cdot{\bf G}(\omega;{\bf r}_{i},{\bf
r}_{j})\cdot{\bf n}_{j}$, with similar decomposition $S_{ij}=S_{ij}^{0}(\omega)+S_{ij}^{g}(\omega)$. In terms of
$S_{ij}$, the transferred power, Eq.~(\ref{power}), takes a simple form $P_{ad}=(\omega/2)\mu_{d}^{2}\alpha''_{a}\left|S_{ad}\right|^{2}$.

ETE is obtained by normalizing $P_{ad}(\omega)$ with respect to the donor's full power loss, $P_{d}(\omega)$, followed by integration over the donor's emission band: $E_{ad}=\int d\omega f_{d}P_{ad}/P_{d}$. In the lowest order, $P_{d}(\omega)$ has the form
\begin{equation}
P_{d}=P_{d}^{0}+P_{d}^{g}+P_{ad},\label{eq:Pd_contbs}
\end{equation}
where $P_{d}^{0}$ stands for the donor's power loss due to radiative and intrinsic (non-radiative) processes and $P_{d}^{g}$ is the power dissipated in graphene. In vacuum, the former is given by $P_{d}^{0}=(\omega/2Q_{d})\mu_{d}^{2}\text{Im}S_{dd}^{0}=\mu_{d}^{2}\omega^{4}/3c^{3}Q_{d}$, where the donor's quantum yield, $Q_{d}$, accounts for intrinsic losses, while in the presence of dielectric interface (formed by an under-graphene substrate, e.g., ${\rm SiO_{2}}$), it is more involved \cite{Chance1978-1} and has been evaluated by us numerically. The power dissipated in graphene is given by $P_{d}^{g}=(\omega/2)\mu_{d}^{2}\text{Im}S_{dd}^{g}$. For $z_{i}>1$ nm considered here, higher order terms describing feedback from acceptor to 
graphene and from graphene to donor \cite{Pustovit2011-085427} are small and, therefore, neglected. The ETE then takes the form
\begin{equation}
E_{ad}=\int d\omega\,\frac{f_{d}\alpha''_{a}\left|S_{ad}\right|^{2}}{\text{Im}S_{dd}^{0}/Q_{d}+\text{Im}S_{dd}^{g}+\alpha''_{a}\left|S_{ad}\right|^{2}},\label{rate}
\end{equation}
where averaging over dipoles' orientations is implied (see Appendix~\ref{app:orient} for the detailed discussion of the ETE dependence on specific dipoles' orientations).

We now proceed with evaluation of $S_{ij}=S_{ij}^{0}(\omega)+S_{ij}^{g}(\omega)$. The direct (F\"orster) ETE is
determined by the Coulomb term in free space, $S_{ad}^{0}=q_{ad}/r_{ad}^{3}$,\footnote{In our numerical calculations,
the effect of dielectric screening by the under-graphene substrate is accounted for exactly through image charges. At
$R_{ad}\gg z_{a,d}$, this reduces to $S_{ad}^{0}=q_{ad}/\tilde{\kappa}R_{ad}^{3}$, which is what is assumed in deriving
Eq.~(\ref{eq:distance_pl}).} where $q_{ad}$ is the orientational factor with average $\langle q_{ad}^{2}\rangle=2/3$, while the donor's radiative losses are described by $\text{Im}S_{dd}^{0}=\frac{2}{3}(\omega/c)^{3}$. The graphene contribution to $S_{ij}$ can be found as follows. In the longwave limit where retardation effects can be neglected, $S_{ad}(\omega)$ reduces to
\begin{equation}
S_{ad}=-\left({\bf n}_{a}\cdot\nabla_{a}\right)\left({\bf n}_{d}\cdot\nabla_{d}\right)U({\bf r}_{a},{\bf r}_{d})\label{longwave}
\end{equation}
where $U=v+v\Pi v$ is the Coulomb potential screened by the graphene sheet, $v({\bf r})=v({\bf R},z)$ is the bare Coulomb potential and $\Pi({\bf R,\omega})$ is the density correlation function of graphene. After the in-plane Fourier transform using $v_{q}(z)=\frac{2\pi}{\tilde{\kappa}q}e^{-q|z|}$, the graphene contribution is obtained as
\begin{equation}
S_{ad}^{g}(\omega)=\frac{e^{2}}{\tilde{\kappa}^{2}}\int d{\bf q}\:g_{a}(\hat{{\bf q}})g_{d}^{*}(\hat{{\bf q}})\Pi(q,\omega)e^{-q|z_{a}|-q|z_{d}|+i{\bf q}\cdot{\bf R}_{ad}},\label{S-graphene}
\end{equation}
where $g_{i}(\hat{{\bf q}})={\bf n}_{i}\cdot\hat{{\bf q}}+i{\bf n}_{i}\cdot\hat{{\bf z}}_{i}$ is the orientational
factor, $\hat{{\bf q}}$ and $\hat{{\bf z}}_{i}$ being, respectively, the radial unit vector in the graphene's plane and the normal from graphene's plane to fluorophore $i$. Eqs~(\ref{rate}) and (\ref{S-graphene}) are used by us to \emph{numerically} evaluate ETE and obtain all the numerical results in this work. Specifically, all the Green dyadics in the matrix representation ($S_{ad}$, $S^0_{dd}$ and $S^g_{dd}$) are first evaluated  (for each $\omega$) via numerical integration over the wavenumber ${\bf q}$ adopting the density correlation function of the homogeneous graphene, $\Pi(q,\omega)$, in the random phase approximation (see Appendix~\ref{app:polariz} for details). Then, the integration over $\omega$ in Eq.~(\ref{rate}) is performed numerically.  

Analytical expression for  the long-distance behavior of $S_{ad}^{g}$ can be derived using the plasmon pole approximation for $\Pi(q,\omega)$ as (see Appendix~\ref{app:polariz}) 
\begin{equation}
\Pi(q,\omega)\approx\Pi^{pp}(q,\omega)=\frac{\Lambda_p}{q_{p}-q+i/2R_{p}},\label{P-pl}
\end{equation}
where $\Lambda_p$, $q_p$ and $R_p$, being respectively the GP amplitude, wavenumber and travel length, are obtained by
locating the resonance of $\Pi(q,\omega)$ at $q=q_p(\omega)$. At low energies (i.e., $\hbar\omega\ll \epsilon_f$), this can be done analytically yielding
\begin{subequations}
\begin{equation}
\Lambda_p=\frac{\tilde{\kappa}q_{p}^{2}}{2\pi e^{2}},
\end{equation}
\begin{equation}
q_{p}=\tilde{\kappa}\hbar^{2}\omega^{2}/2\epsilon_{f}e^{2},
\end{equation}
\begin{equation}
R_{p}=\epsilon_{f}e^{2}/\tilde{\kappa}\hbar^{2}\omega\gamma.\label{eq:Rp_analyt}
\end{equation}
\end{subequations}
It turns out, that even at $\hbar\omega\sim\epsilon_f$ (i.e., regime considered in this work) the low-$\omega$ analytical expressions for $\Lambda_p$ and $q_p$ are still applicable. In contrast, $R_p$ has to be found very accurately (i.e., numerically) since Eq.~(\ref{eq:distance_pl}) is exponentially sensitive to its value. Our additional numerical tests (not shown) have demonstrated that the analytical expression for $R_p$ becomes quite accurate already at $\hbar\omega/\epsilon_f\lesssim$~0.3 -- 0.5. However, at $\hbar\omega\approx\epsilon_f$ one can expect the magnitude of the inaccuracy of $R_p$, as obtained from Eq.~(\ref{eq:Rp_analyt}), to be of the order of the value of $R_p$. Specifically, for the parameters adopted in this paper, the analytically found $R_p$ is approximately twice as high as its numerical counterpart. In what follows, the analytical expressions for $\Lambda_p$, $q_p$, and numerically found $R_p$ are adopted.

Substitution of Eq.~(\ref{P-pl}) into Eq.~(\ref{S-graphene}) yields
\begin{equation}
S_{ad}^{g}=\frac{q_{p}^{2}}{2\pi\tilde{\kappa}}\,\int dqq\frac{e^{-q(|z_{a}|+|z_{d}|)}}{q_{p}-q+i/2R_{p}}\int d\phi f_{a}(\hat{{\bf q}})f_{d}^{*}(\hat{{\bf q}})e^{i{\bf q}\cdot{\bf R}},\label{S-pl}
\end{equation}
where $\phi$ is the azimuthal angle in ${\bf q}$-plane. For $qR\gg1$,
only small fluctuations of $\phi$ around ${\bf q}\cdot{\bf R}=\pm qR$
contribute to $\phi$-integral $I$, and in these regions $f_{i}(\hat{{\bf q}})$
can be replaced by $f_{i}(\pm\hat{{\bf R}})$, yielding
\begin{equation}
I=\left(\frac{8\pi}{qR}\right)^{1/2}\text{Re}\left[e^{iqR-i\pi/4}f_{a}(\hat{{\bf R}})f_{d}^{*}(\hat{{\bf R}})\right].\label{int}
\end{equation}
Upon substitution of Eq.~(\ref{int}) into Eq.~(\ref{S-pl}), $S_{ad}^{g}$
splits into two parts corresponding to the outgoing and incoming waves,
$e^{\pm iqR}$. For $qR\gg1$, the dominant contribution comes from
the pole at $q=q_{p}+i/2R_{p}$ into the outgoing part. Finally, after
averaging of $\left|S_{ad}^{g}\right|^{2}$ over dipoles' orientations
using relations $\langle f_{i}f_{j}^{*}\rangle=\frac{2}{3}\delta_{ij}$
and $\langle f_{i}f_{j}\rangle=0$, one obtains
\begin{equation}
\left|S_{ad}^{g}\right|^{2}=\frac{8\pi q_{p}^{5}}{9\tilde{\kappa}^{2}R_{ad}}\, e^{-R_{ad}/R_{p}-2q_{p}(|z_{a}|+|z_{d}|)}.\label{matrix-pl}
\end{equation}
Comparison of Eq.~(\ref{matrix-pl}) and direct contribution $\left|S_{ad}^{0}\right|^{2}=\frac{2}{3}r_{ad}^{-6}$ reveals
that the GP-assisted ET channel is dominant for $R_{ad}\gtrsim\lambda_{p}$. Specifically,  numerical calculations point
to a crossover to the GP-assisted regime at $R_{ad}\sim30$ -- $40$ nm for $\lambda_{p}\approx 30$ nm (see Figure~\ref{fig:Ead_num_anl}). GP-assisted ET is ineffective for large fluorophores' distances to the graphene plane ($z_{i}q_{p}\gg1$) or for their in-plane separation significantly exceeding plasmon travel length ($R_{ad}/R_{p}\gg1$).

Turning to dissipated power in graphene, $P_{d}^{g}$, the diagonal element $S_{dd}^{g}$ can be obtained from Eqs.~(\ref{S-graphene}) and (\ref{P-pl}) via substitution $a\rightarrow d$, which yields
$\text{Im}S_{dd}^{g}=\frac{2\pi}{3}\frac{q_{p}^{3}}{\tilde{\kappa}}e^{-2q_{p}|z_{d}|}$.\cite{Velizhanin2011-085401}
If the acceptor is absent, $P_{d}^{g}$ determines the normalized energy loss rate of the donor, $k_{d}=(P_{d}^{0}+P_{d}^{g})/P_{d}^{0}$,
shown in Figure~\ref{fig:schematic}(c) for several values of $\hbar\omega$. At very large $z_d$, the donor does not ``feel" the presence of graphene, so its losses are dominated by radiative and non-radiative ones. At smaller distances, the exponential decay of $k_{d}$ with the donor-graphene distance indicates the predominant donor's energy transfer to GP. At even smaller fluorophore-graphene distances, the non-exponential dependence of $k_{d}$ on $z_d$ is due to the onset of excitations of electron-hole pairs in graphene [see inset in Figure~\ref{fig:schematic}(c)]. These three regimes of a single fluorophore interaction with graphene have recently been studied in detail elsewhere.\cite{Velizhanin2011-085401,Koppens2011-3370,Huidobro2012-155438}

The above considerations lead to a conclusion that in the wide range of intermediate donor-graphene distances (i) donor's energy losses are dominated by GP excitation, and (ii) $k_{d}\gg 1$ and so  $P_{d}^{g}$ dominates over intrinsic and radiative losses. Furthermore, at distances between fluorophores exceeding F\"{o}rster radius, $P^g_{d}$ dominates over $P_{ad}$ in Eq.~(\ref{eq:Pd_contbs}) as well. Thus, in a wide parameter range, both the numerator and denominator of the integrand in Eq.~(\ref{rate}) are dominated by GP-assisted channels, yielding Eqs.~(\ref{eq:ete_pl}) and (\ref{eq:distance_pl}) for ETE.

\section{Results and Discussion}\label{sec:RD}

ET calculations below were performed for a pair (donor and acceptor) of PbSe QDs with emission and absorption bands centered at  0.55 eV and 0.6 eV, respectively  \cite{Pietryga2004-11752,Nemova2012-676}. The fluorescence quantum yield for such QDs varies significantly in literature, \cite{Pietryga2004-11752,Liu2010-14860} so the ``average'' value of $10^{-2}$ is adopted here. Lorentzian lineshape for both bands is assumed with full width at half maximum (FWHM) of 0.1 eV,\cite{Steckel2003-1862,Pietryga2004-11752,Nemova2012-676} and the acceptor absorption crossection is chosen $\sigma_{a}=\left (4\pi\omega/3c\right )\alpha''=2$~\AA$^{2}$ at its spectral maximum.\cite{Brumer2006-7488,Cheng2008-1404,Nemova2012-676}  Both optical bands lie within the GP band with dispersion $\omega\propto\sqrt{q}$ in doped graphene with electron scattering rate chosen as $\gamma=10$~ps$^{-1}$.\cite{Jablan2009-245435,Koppens2011-3370} For the Fermi level at $\epsilon_{f}=0.6$ eV adopted here,\cite{Chen2011-617} GP is well defined up to $q\approx 0.6$ nm$^{-1}$ corresponding to $\hbar\omega\approx 0.8$ eV, while for larger $q$ GP is dampened by interband single-particle transitions -- Landau damping [see Figure~\ref{fig:schematic}(b)]. 
A donor with emission band centered at 0.55 eV [dashed horizontal line in Figure~\ref{fig:schematic}(b)] predominantly excites GPs with $q_p\approx 0.2$ nm$^{-1}$  (red dagger), while excitation of electron-hole pairs requires higher wavenumbers $q\gtrsim 0.9$ nm$^{-1}$ (blue half-oval) and is, therefore, efficient only for $z_{d}<1$ nm [see inset in Figure~\ref{fig:schematic}(c)]. Below we choose the values $z_{d}=z_{a}=3$ nm lying in the GP-dominated exponential domain with $k_{d}\approx 5\times 10^{3}$ [see Figure~\ref{fig:schematic}(c)]. 

In Figure~\ref{fig:Ead_num_anl}, the results of our numerical and analytical calculations of ETE for suspended graphene are compared to F\"{o}rster's ETE for a similar system in vacuum. F\"{o}rster's ETE shows characteristic behavior described by standard expression  $E_{ad}^{F}=\left (1+r_{ad}^{6}/r_F^{6}\right )^{-1}$, where calculated F\"{o}rster radius $r_{F}\approx8$~\AA~is consistent with experimental results for a similar system.\cite{Clark2007-7302} Numerical results for ETE in the presence of graphene (solid lines) are shown for several values of electron scattering rate $\gamma$. For small $R_{ad}$, the energy {\em transfer} from donor to acceptor, determined by the integrand's numerator in Eq.~(\ref{rate}), is dominated by the direct F\"{o}rster mechanism. However, the donor energy {\em losses}, defined by the integrand's denominator, are greatly increased, as compared to the vacuum case, due to the presence of graphene. Under these conditions, the ETE dependence on the distance between QDs becomes F\"{o}rster-like again with $E_{ad}=\left (1+r_{ad}^{6}/r_g^{6}\right )^{-1}$. However, the effective transfer radius, $r_g\approx 2$~nm, is now significantly smaller than $r_F$ in the vacuum case due to a much larger, compared to radiative and intrinsic losses, donor energy dissipation to graphene. 

For large $R_{ad}$, ETE exhibits significant dispersion for different values of $\gamma$ caused by reduction of plasmon travel length $R_{p}$ with increasing $\gamma$ [see Eq.~(\ref{P-pl})] and, hence, the exponential suppression of ETE for $R_{ad}\gtrsim R_{p}$. The distance dependence of ETE for $R_{ad}\gtrsim 10$~nm is in excellent agreement with our analytical results, Eqs.~(\ref{eq:ete_pl}) and (\ref{eq:distance_pl}), shown by symbols in Figure~\ref{fig:Ead_num_anl}.

\begin{figure}
\includegraphics[width=1.0\columnwidth]{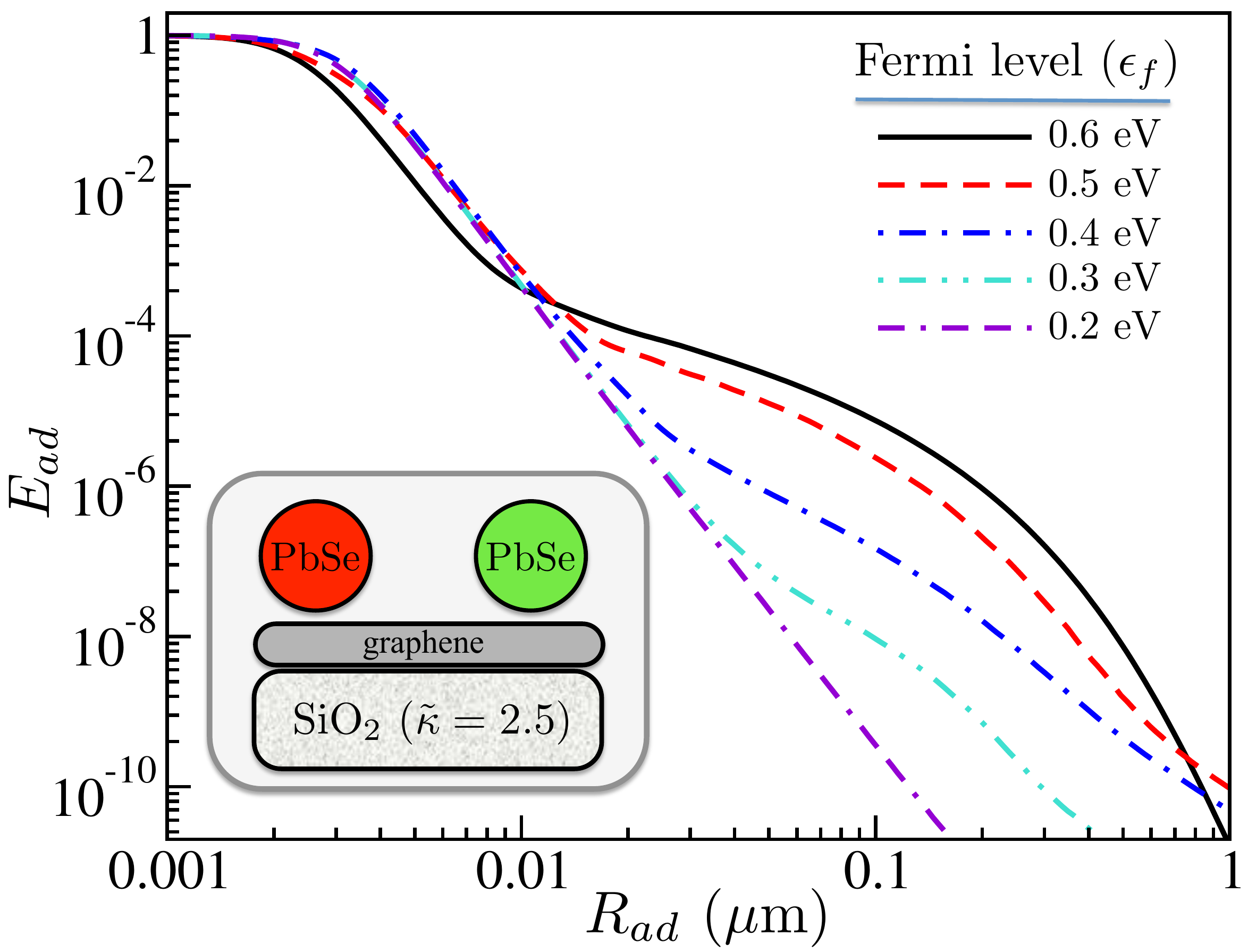} \caption{\label{fig:Ead_ef}
Dependence of ETE between two PbSe QDs situated near graphene on the doping level of graphene. The adopted parameters are $\gamma=10$~ps$^{-1}$ and $z_d=z_a=3$~nm. The schematic of the system is shown in the inset.}
\end{figure}

In Figure~\ref{fig:Ead_ef} we show how ETE evolves with the doping level of graphene.
As the Fermi level is reduced from 0.6 eV to 0.2 eV with the decrement of 0.1 eV, ETE first decreases slowly and then sharply drops at $\epsilon_{f}$ below 0.5 eV, i.e., when GP Landau damping onset ($\approx$1.3$\epsilon_f$) moves below donor's emission band. For $\epsilon_f=0.2$ eV, ETE shows F\"{o}rster-like behavior $\propto$$R_{ad}^{-6}$ but with the reduced effective radius of $\approx$2.5 nm due to ET quenching by graphene.\cite{Swathi2008-054703,Swathi2009-086101,Velizhanin2011-085401}

\begin{figure}
\includegraphics[width=1.0\columnwidth]{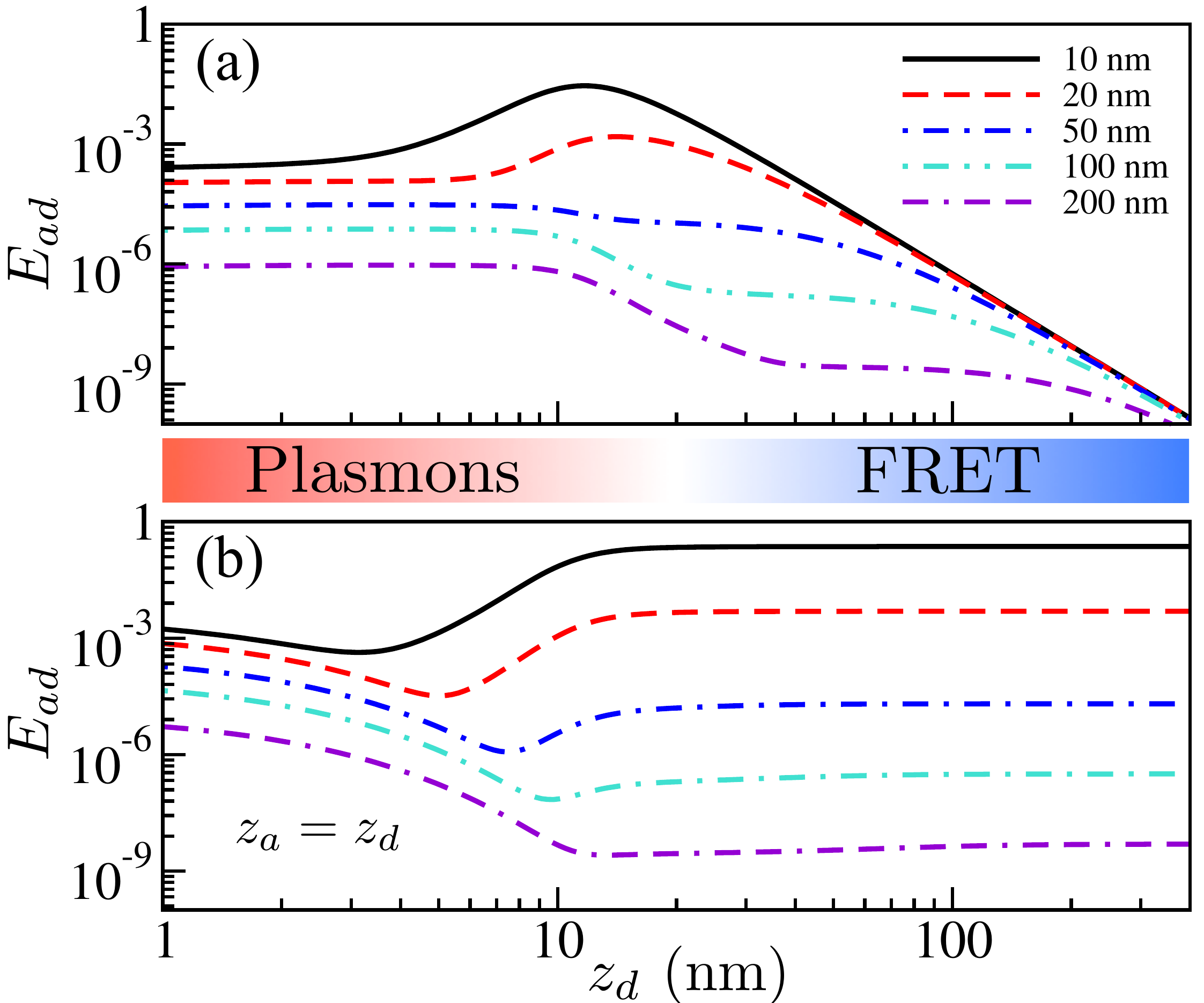} \caption{\label{fig:Ead_z}
Dependence of ETE on the distance between donor/acceptor and graphene. Legend encodes $R_{ad}$. (a) Donor-graphene distance, $z_d$, is varied, while the acceptor-graphene distance kept constant ($z_{a}=3$ nm). (b) Both distances are varied simultaneously ($z_{a}=z_{d}$). Doped graphene ($\epsilon_{f}=0.6$ eV) lays on top of the ${\rm SiO_{2}}$ substrate ($\tilde{\kappa}=2.5$). The adopted electron scattering rate is $\gamma=10$~ps$^{-1}$. The gradient-colored bar between the panels marks the transition from the GP-mediated ET at small $z_d$ to the standard FRET mechanism at larger $z_d$.}
\end{figure}

In Figure~\ref{fig:Ead_z}, we plot calculated ETE vs. fluorophores' separation from the graphene sheet  ($z_{a},z_{d}$) for different values of in-plane distance $R_{ad}$.
Here, we distinguish between two scenarios: (a) the acceptor's position is fixed, whereas the donor is moved away from the graphene plane, so that donor-acceptor distance $r_{ad}$ increases as well; and (b) the acceptor follows the donor so that both fluorophores are moved \textit{in sinc} away from the graphene plane, i.e., $r_{ad}$ stays constant for each $R_{ad}$ value. In both scenarios, at $z_d\lesssim 10$ -- $20$~nm both ET and losses are dominated by the GP-assisted channel, and, therefore, Eqs.~(\ref{eq:ete_pl}) and (\ref{eq:distance_pl}) are expected to provide an accurate description of ETE behavior. Indeed, ETE plateaus [panel (a)] and the exponential decay of ETE [panel (b)] at low $z_d$ both originate from the same exponent in Eq.~(\ref{eq:distance_pl}), which is independent of $z_d$ and linear with respect $z_a$, respectively. 

At large $z_d$, the GP amplitude is exponentially dampened [Eq.~(\ref{S-graphene})], i.e., graphene becomes effectively absent from the ET picture, so that ETE dependence on $z_d$ follows the standard FRET-like $r^{-6}_{ad}$ behavior. Specifically, $E_{ad}\propto R^{-6}_{ad}$ behavior for $z_a=z_d$ results in plateaus with $R_{ad}$-dependent levels at large $z_d$ [panel (b)]. In panel (a), this $r^{-6}_{ad}$ dependence reduces to $z_d^{-6}$ at very large $z_d$ (i.e., $z_d\gg R_{ad},z_a$). If the GP-assisted channel is already negligible but $z_d$ is still much smaller than $R_{ad}$ -- this regime can be realized at $R_{ad}\gg \lambda_p\approx 30$~nm -- then $r_{ad}\approx R_{ad}$ and $E_{ad}$ levels off with respect to $z_d$ at, e.g.,  $z_d=30$ -- $100$~nm for $R_{ad}=100$ and 200~nm [panel (a)]. Note that the magnitudes of large-$z_d$ plateaus in panels (a) and (b) match for each $R_{ad}$ value.

Finally, Figure~\ref{fig:Ead_z}(a) shows that the transition from GP-dominated to FRET-dominated ET results in the ETE increase for $R_{ad}=10$ -- $20$~nm and its decrease for larger in-plane distances. Bearing in mind the effective ``absence" of graphene at large $z_d$, this behavior can be traced back to that in Figure~\ref{fig:Ead_num_anl}, where ET without graphene (i.e., FRET) is more efficient than the GP-mediated ET in the presence of graphene at $R_{ad}\lesssim 30$~nm,  and less efficient for larger in-plane distances.  

\section{Conclusion}

In this paper, we have shown that a single-atom layer of doped graphene can be used for highly efficient long-range energy transfer at the nanoscale. The transfer is mediated by longitudinal plasmons in graphene and hence it is very sensitive to the sample mobility and doping level which determine plasmon lifetime and travel length. We have demonstrated that in clean samples with high doping levels (e.g., $\epsilon_{f}=0.6$ eV), the energy transfer efficiency can exceed that of FRET by up to $\sim$$10^{4}$ at hundreds nm distances. For a given donor-acceptor pair, the transfer efficiency can be optimized by tuning parameters of the system, e.g., fluorophore-graphene distances.

\acknowledgements

Work at LANL was performed under the NNSA of
the U.S. DOE at LANL under Contract No. DE-AC52-06NA25396. Work at JSU was supported by the NSF under Grants No. DMR-1206975 and No. HRD-0833178 and under EPSCOR program.

\appendix

\section{Density correlation function}\label{app:polariz}

The bare density correlation function, or retarded polarization operator,
is calculated within the Dirac electrons approximation as \cite{Wunsch2006-318,Hwang2007-205418,Koppens2011-3370}
\begin{align}
\Pi_{0}(q,\omega)= & \frac{1}{4\pi\hbar}\left[\frac{8\epsilon_{f}}{\hbar v_{f}^{2}q^{2}}+\frac{G(-\Delta_{-})\theta\left[-{\rm Re}\left\{ \Delta_{-}\right\} -1\right]}{\sqrt{\omega^{2}-v_{f}^{2}q^{2}}}\right.,\nonumber \\
+ & \left.\frac{\left[G(\Delta_{-})+i\pi\right]\theta\left[{\rm Re}\left\{ \Delta_{-}\right\} +1\right]-G(\Delta_{+})}{\sqrt{\omega^{2}-v_{f}^{2}q^{2}}}\right],\label{eq:full_pop}
\end{align}
where $G(z)=z\sqrt{z^{2}-1}-\ln\left(z+\sqrt{z^{2}-1}\right)$ and
$\Delta_{\pm}=\left(\omega/v_{f}\pm2\epsilon_{f}/\hbar v_{f}\right)/q$.
The square roots are chosen to yield positive real parts and the imaginary
part of the logarithm is taken in $(-\pi,\pi]$ range. Fermi velocity and Fermi level (the latter determines the extent of graphene doping) are denoted by $v_f$ and $\epsilon_f$, respectively. Within the Dirac electrons approximation, the density correlation function  is insensitive to the sign of the Fermi level, so in all the expressions here and in the main text $\epsilon_f$ has to be understood as $|\epsilon_f|$.

The two important limiting forms of the density correlation function are (i) the long wavelength limit ($q\rightarrow0$, $\hbar\omega\ll2\epsilon_{f}$),  and (ii) the static limit ($\omega\rightarrow 0$, $q<2k_f$). The long wavelength limit is given by
\begin{equation}
\Pi_{0}(q\rightarrow 0,\omega)=\frac{\epsilon_{f}q^{2}}{\pi\hbar^{2}\omega^{2}}.\label{eq:pop_lowq}
\end{equation}
The static limit of the bare density correlation function is obtained as
\begin{equation}
\Pi_{0}(q,\omega\rightarrow 0)=-\frac{2\epsilon_{f}}{\pi\hbar^{2}v^2_f}.\label{eq:pop_stat}
\end{equation}
 
The naive substitution $\omega\rightarrow\omega+i\gamma/2$ to account
for in-graphene scattering losses in Eq.~(\ref{eq:full_pop}) ($\gamma$ is the electron scattering
rate) is inaccurate in a general case (especially if $\gamma$
is not small), since it does not preserve the particle conservation requirement.
To correct for this, the more accurate Mermin procedure is adopted,
yielding \cite{Mermin1970-2362,Ropke1999-365}
\begin{equation}
\Pi_{\gamma}(q,\omega)=\frac{(1+i\gamma/\omega)\Pi_{0}(q,\omega+i\gamma)}{1+(i\gamma/\omega)\Pi_{0}(q,\omega+i\gamma)/\Pi_{0}(q,0)}. \label{eq:Mermin}
\end{equation}

The full (or ``dressed") density correlation function, which accounts for screening
in graphene, is obtained within the random phase approximation as
\begin{equation}
\Pi(q,\omega)=\frac{\Pi_{\gamma}(q,\omega)}{1-e^{2}v(q)\Pi_{\gamma}(q,\omega)},\label{eq:pop_rpa}
\end{equation}
where $v(q)=2\pi/\tilde{\kappa}q$ is the two-dimensional Fourier transform of the
Coulomb potential within the graphene's plane, $v(R)=1/\tilde{\kappa}R$. The effective dielectric
constant of the environment is given by $\tilde{\kappa}=(\kappa_{1}+\kappa_{2})/2$
for a graphene sheet sandwiched between two homogeneous dielectrics
with dielectric constants $\kappa_{1}$ and $\kappa_{2}$.\cite{Smythe1968,Ponomarenko2009-206603}
Thus $\tilde{\kappa}=1$ for a suspended graphene sheet in vacuum.
For graphene, laid on top of a ${\rm SiO_{2}}$ substrate ($\kappa_{1}=1$,
$\kappa_{2}=\kappa_{{\rm SiO_{2}}}=4$), one obtains $\tilde{\kappa}=2.5$.

The plasmon dispersion relation, $q_{p}=q_{p}(\omega)$, is found by requiring the real part of the denominator of Eq.~(\ref{eq:pop_rpa}) to vanish.  The Taylor expansion of the denominator around this point (up to leading terms in both real and imaginary parts) leads to the possibility of approximating  the full density correlation function within the so called plasmon pole approximation as
\begin{equation}
\Pi^{pp}(q,\omega)=\frac{\Lambda_p}{q_{p}-q+i/2R_p},\label{eq:plpole}
\end{equation}
where $\Lambda_p=\Pi_\gamma(q_p,\omega)/A$ is the plasmon amplitude, and $R_p=A/2B$ is the plasmon travel length. The coefficients of the Taylor expansion of the denominator of Eq.~(\ref{eq:pop_rpa}) are
\begin{align}
A&=e^2\left.\frac{\partial}{\partial q}\left(v(q){\rm Re}\left[\Pi_{\gamma}(q,\omega)\right]
\right)\right|_{q=q_{p}},\nonumber \\
B&=-e^{2}v(q_{p}){\rm Im}\left[\Pi_{\gamma}(q_{p},\omega)\right].
\end{align}
In the low-$\omega$ limit (i.e., $\hbar\omega\ll\epsilon_f$), the density correlation function in the plasmon pole approximation can be obtained purely analytically by (i) substituting Eqs.~(\ref{eq:pop_lowq}) and (\ref{eq:pop_stat}) into Eq.~(\ref{eq:Mermin}), and (ii) using the so obtained $\Pi_\gamma(q,\omega)$ to evaluate the Taylor expansion coefficients $A$ and $B$. The result of these manipulations -- the explicit analytical formulas for $\Lambda_p$, $q_p$
and $R_p$ -- is provided in the main text after Eq.~(\ref{P-pl}). 

It turns out that for the specific case considered here, i.e., the plasmon pole approximation in the long wavelength limit, the same analytical expression for $\Pi^{pp}(q,\omega)$ could have been obtained in the limit of small $\gamma$ by using the substitution $\omega\rightarrow\omega+i\gamma/2$ instead of the more general Mermin's procedure. It has to be emphasized, however, that such an agreement is not general and hard to foresee. Therefore, the more accurate Mermin's procedure has to be favored over more approximate methods of introducing the finite scattering rate into the density correlation function.\cite{Ropke1999-365}

\section{Dipoles' orientation dependence of ETE}\label{app:orient}

\begin{figure}
\includegraphics[width=1.0\columnwidth]{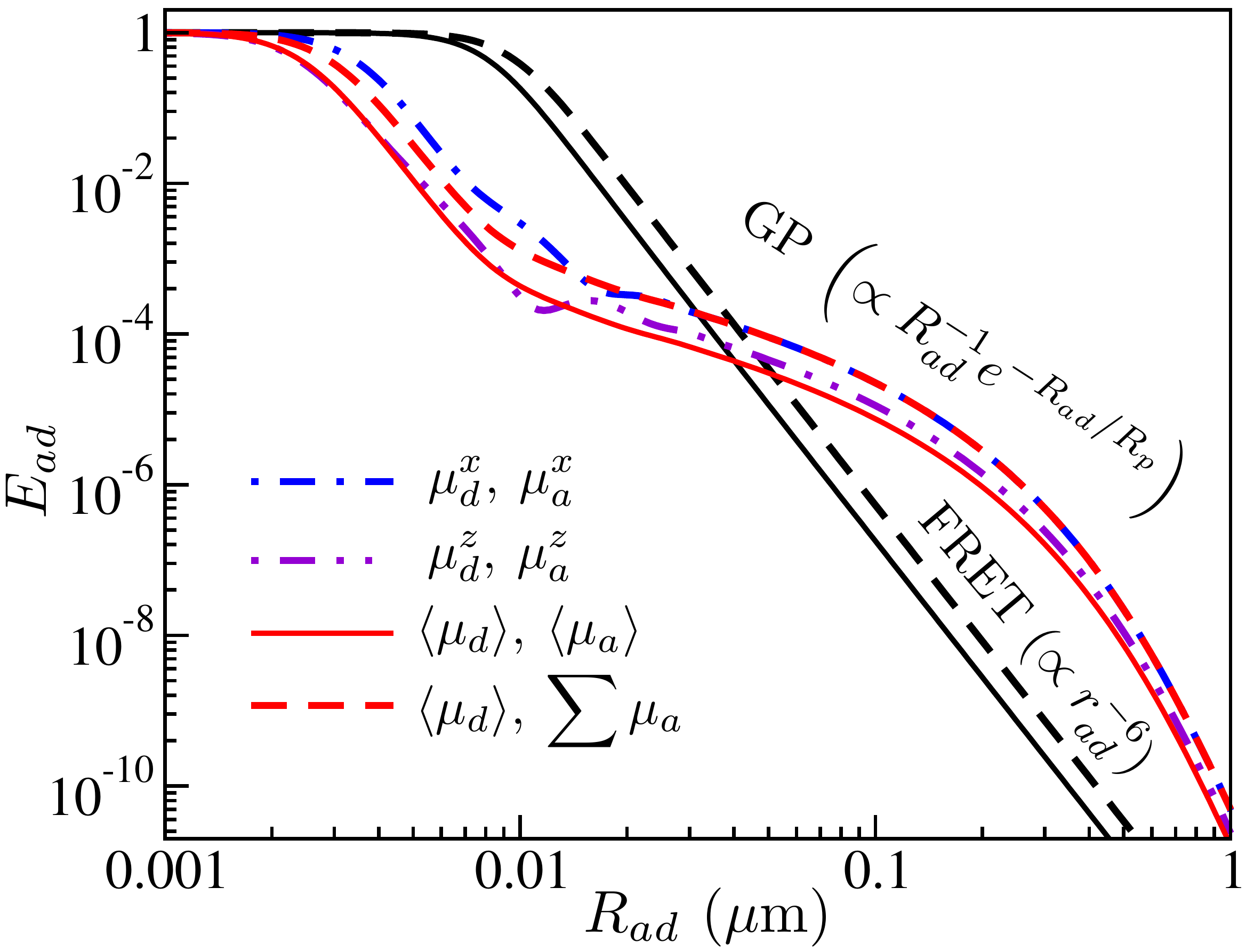} \caption{\label{fig:Ead_prjs}
ETE without (black) and
with graphene (color lines) laid on top of the ${\rm SiO_{2}}$
substrate ($\tilde{\kappa}\approx2.5$). The adopted parameters are $\epsilon_f=0.6$~eV, $\gamma=10$~ps$^{-1}$ and $z_a=z_d=$3~nm. Dash-dotted, dash-double-dotted, and
solid lines correspond to cases where both the donor and acceptor
transition dipole vectors are elongated in $x$- and $z$-directions,
or averaged over all the directions, respectively. Dashed lines
represent \emph{summing} over the acceptor's dipole projections instead
of averaging.}
\end{figure}

Figure~\ref{fig:Ead_prjs} shows the distance dependence of ETE for fluorophores near graphene at different dipole orientations. 
Graphene is laid on top of the ${\rm SiO_{2}}$ substrate ($\tilde{\kappa}=2.5$) and the Fermi level is set to $\epsilon_f=0.6$~eV. The electron scattering rate is assumed $\gamma=10$~ps$^{-1}$.

Dash-dotted and dash-double-dotted lines represent the both donor and acceptor transition dipoles
fixed in the $x$-direction [schematically shown in Figure~\ref{fig:schematic}(a) in the main text]
and in the $z$-direction (dipoles perpendicular to the graphene's plane),
respectively. ETE in the projection-averaged case is shown by solid lines. 

Dashed lines represent the case where the \emph{summation}
is performed over the acceptor's dipole projections instead of averaging (solid lines). This has to be done if the acceptor's dipole polarizability is isotropic, which is frequently the case for spherical semiconductor QDs as fluorophores. For example, this is true for PbSe QDs, where the dipole polarizability is isotropic due to the presence of four degenerate $L$-valleys corresponding to the four equivalent $\langle 111\rangle$ directions in the face-centered cubic lattice of lead chalcogenides.\cite{Kang1997-1632}
Within the analytical plasmon pole approximation, to substitute the averaging over acceptor's projections with summation it suffices to add an extra factor of 3 into Eq.~(\ref{matrix-pl}) in the main text.

For comparison, the dependence of the F\"orster ETE (i.e., in the absence of graphene) on dipole orientations is shown by black lines. In this case, the standard F\"orster ETE with $E_{ad}=(1+r_{ad}^{6}/r_{F}^{6})^{-1}$
is recovered. The slightly
smaller F\"orster radius, $r_{F}\approx7.5$ nm for the solid black line,
than the one obtained for the suspended graphene ($r_{F}\approx8$
nm) (Figure~\ref{fig:Ead_num_anl} in the main text), is due to the SiO$_2$-induced dielectric screening ($\tilde{\kappa}=2.5$).

Specific dipole orientations can lead to strong ETE variations in the crossover region between F\"{o}rster-dominated and GP-dominated regimes. This behavior is due to the interference between the F\"orster and GP contributions to ET in the region where the magnitudes of these two contributions are comparable. In particular, the negative and positive interferences are seen for dipole projections fixed in $z$ and $x$-directions, respectively, at $R_{ad}\approx 10$~nm in Figure~\ref{fig:Ead_prjs}.

At large donor-acceptor distances, a specific dipole orientation has no significant effect on ETE, except for the overall numerical factor of the order of $\sim$1. For example, the donor with the transition dipole fixed in $z$-direction is twice as efficient in exciting GP than that with the dipole in $x$-direction.\cite{Velizhanin2011-085401} However, the $z$-dipole excites plasmons isotropically within the graphene plane, whereas the GP emission of $x$-dipole has a characteristic dipolar pattern [see Figure~\ref{fig:schematic}(a) in the main text], {\em concentrated} in the direction of the acceptor (and also in the opposite direction). This results in the {\em same} power transfer in both cases, but with {\em lower} power losses in the case of the dipole fixed in the $x$-direction, which ultimately yields twice as high ETE for $x$-dipoles than for $z$-dipoles. 


%

\end{document}